\documentclass[twocolumn,superscriptaddress,longbibliography,aps,
    prb,preprintnumbers,floatfix,10pt]{revtex4-2}

\usepackage{graphicx}
\usepackage{bm}
\usepackage{color}
\usepackage{epstopdf}
\usepackage{amsmath}
\usepackage{amssymb}
\usepackage{epstopdf}
\usepackage{gensymb}

\usepackage{float}

\usepackage[urlcolor=blue,colorlinks=true,citecolor=blue,linkcolor=blue,pdfstartview={FitH},bookmarks=false]{hyperref}

\graphicspath{{figs/}{./fig/}{.}}

\begin{document}

\title{Electronic and magnetic properties of the NdNiO$_2$/SrTiO$_3$ thin films}

\author{Sajid Sekh}
\affiliation{Institute of Nuclear Physics, Polish Academy of Sciences,  
ul. W. E. Radzikowskiego 152, PL-31342 Krak\'{o}w, Poland} 

\author{Andrzej Ptok}
\email[e-mail: ]{aptok@mmj.pl}
\affiliation{Institute of Nuclear Physics, Polish Academy of Sciences,  
ul. W. E. Radzikowskiego 152, PL-31342 Krak\'{o}w, Poland} 

\author{Wojciech Brzezicki}
\affiliation{Institute of Theoretical Physics, Jagiellonian University,
Prof. Stanis\l{}awa \L{}ojasiewicza 11, PL-30348 Krak\'{o}w, Poland}

\author{Przemys\l{}aw Piekarz}
\email[e-mail: ]{przemyslaw.piekarz@ifj.edu.pl}
\affiliation{Institute of Nuclear Physics, Polish Academy of Sciences,  
ul. W. E. Radzikowskiego 152, PL-31342 Krak\'{o}w, Poland} 

\author{Andrzej~M.~Ole\'{s}}
\affiliation{Institute of Theoretical Physics, Jagiellonian University,
Prof. Stanis\l{}awa \L{}ojasiewicza 11, PL-30348 Krak\'{o}w, Poland}

\begin{abstract}
The hole-doped NdNiO$_2$ layer deposited on the SrTiO$_{3}$ surface exhibits unconventional superconductivity.
Here, we present a systematic study of the electronic and magnetic properties of the NdNiO$_2$ superconductor using the density functional theory (DFT). 
The strong local Coulomb interactions in the Ni($3d)$ and Nd($4f$) states are included within the DFT+$U$ method. 
The effect of Sr doping on the electronic band structure and density of states was studied for the NdNiO$_2$ thin films deposited on the SrTiO$_3$ (001) surface.
The results obtained for the uncapped thin films were compared with the 
calculations for the NdNiO$_2$ films capped by the SrTiO$_3$ layer.
We have found significant changes in the electronic structure and magnetic
properties of the thin films compared to the bulk crystal.
\end{abstract}

\maketitle

\section{Introduction}

Recently, unconventional superconductivity was reported in hole-doped infinite-layer nickelate  
NdNiO$_{2}$ at $9$-$15$~K, in thin-film samples grown on SrTiO$_{3}$~\cite{li.lee.19,zeng.tang.20,gu.li.20}.
Contrary to the thin layers, the bulk NdNiO$_{2}$ does not exhibit the same superconductivity~\cite{wang.zheng.20}.
Superconductivity was also reported in the hole-doped PrNiO$_{2}$ thin films~\cite{osada.wang.20,osada.wang.20b,wang.yang.22,ren.li.23}, while its occurrence in LaNiO$_{2}$ is still under debate~\cite{li.lee.19,osada.wang.21}.
The pressure-induced superconductivity was observed in the bilayer La$_3$Ni$_2$O$_7$ compound at $T_c=80$~K~\cite{sun.huo.23}, as well
as in the trilayer La$_4$Ni$_3$O$_{10}$ material around 25~K~\cite{zhang.pei.25}.
Also under high pressure, superconductivity was found in the monolayer-trilayer phase of La$_3$Ni$_2$O$_7$~\cite{huang.li.25},
which was suggested in Ref.~\cite{abadi.xu.25}.

These observations renewed interest in the pairing mechanism, and the role played 
by the $f$ electrons in rare-earth nickelates~\cite{nomura.arita.22}. 
It was established recently that depending on the interactions in a two-band 
model for infinite-layer (IL) superconductors~\cite{Adh20}, pairing with 
$s$-wave or $d$-wave symmetry is possible~\cite{Pli22}. In this context, numerous theoretical investigations of NdNiO$_2$ have been carried out within density-functional theory (DFT) frameworks~\cite{nomura.hirayama.19,jiang.si.19,Cho20,zhang.jin.20,wu.disante.20,zhang.lane.21,sakakibara.usui.20,botana.norman.20}. Electron–electron correlations were treated either at the static mean-field level using the DFT+$U$ approach~\cite{botana.norman.20,liu.ren.20,wan.ivanov.21,lang.jiang.21,deng.jiang.21,gu.zhu.20}, or more comprehensively by incorporating dynamical mean-field theory (DMFT) within the DFT+DMFT formalism~\cite{wang.kang.20,gu.zhu.20,lechermann.20,ryee.yoon.20,leonov.skornyakov.20,lechermann.20b,karp.botana.20,kitatani.si.20,lechermann.21,liu.xu.21,karp.hampel.22,chen.jiang.22}.

The long-range magnetic order has not been observed experimentally in the accessible temperature range. Nevertheless, recent experiments have reported the presence of magnetic correlations in LaNiO$_2$~\cite{ortiz.puphal.22}, as well as magnetic and charge instabilities in NdNiO$_2$~\cite{lu.rossi.21,zhou.zhang.22,tam.choi.22,krieger.martinelli.22,fowlie.hadjimichael.22}. The charge density wave (CDW) observed in NdNiO$_2$, characterized by the same in-plane wavevector $(1/3,0)$ for Nd ($5d$) and Ni ($3d$) orbitals, disappears upon the emergence of superconductivity in doped NdNiO$_2$~\cite{tam.choi.22}. Similarly, in LaNiO$_2$ the CDW is associated with an incommensurate wavevector; upon doping, the charge order (CO) is progressively suppressed and its wavevector shifts toward a commensurate value~\cite{rossi.osada.22}. Recent DFT studies indicate that antiferromagnetic (AFM) interactions can drive the CDW order in undoped IL nickelates~\cite{zhang.cai.24}. These findings suggest the presence of charge order and its possible interplay with AFM fluctuations and superconductivity in IL nickelates.

\begin{figure}[]
    \centering
    \includegraphics[width=\linewidth]{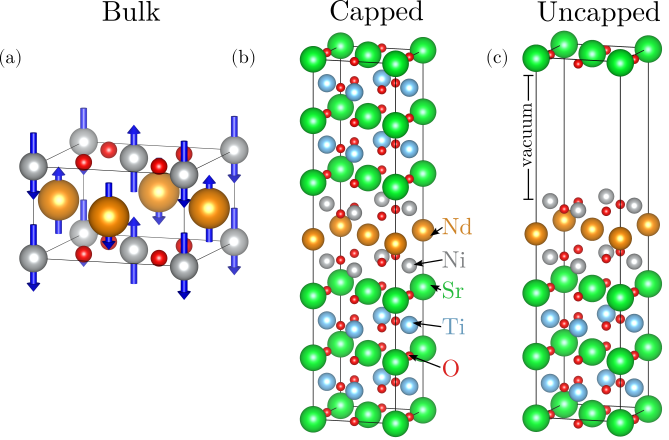} 
    \caption{The tetragonal structures of (a) bulk $\mathrm{NdNiO_2}$ superconductor with the C-AFM order, 
    (b)~capped $\mathrm{NdNiO_2/SrTiO_3}$ (001) interface, and (c) uncapped 
    $\mathrm{NdNiO_2/SrTiO_3}$ (001) interface are shown.}
    \label{fig.cell}
\end{figure}

The measurements for the La$_{1-x}$Sr$_x$NiO$_2$/SrTiO$_3$ thin films demonstrated that $f$ electrons are not essential for superconductivity in IL nickelates~\cite{osada.wang.21}. However, the electron energy-loss spectroscopy measurements provide direct evidence for the multiband electronic structure and the effects of hole doping on the oxygen, nickel and rare-earth bands~\cite{berit.danfend.21}.

The absence of superconductivity in bulk NdNiO$_2$ highlights the crucial role of interface-related effects such as strain, as well as structural and electronic reconstructions in the thin-film heterostructures~\cite{li.he.20,wang.zhang.20,zhou.qin.22}. Previous DFT studies have revealed a pronounced influence of polar surfaces and interfaces on the electronic structure of NdNiO$_2$/SrTiO$_3$ thin films~\cite{geisler.pentcheva.20,he.jiang.20,zhang.lin.20}. At the interface, a two-dimensional electron gas associated with the Ti $3d$ states may emerge~\cite{geisler.pentcheva.20}, while polar instabilities drive lattice reconstructions~\cite{he.jiang.20,zhang.lin.20}. Atomic-resolution electron energy-loss spectroscopy has further demonstrated the formation of a single intermediate Nd(Ti,Ni)O interfacial layer~\cite{goodge.geisler.23}. In addition, studies of NdNiO$_2$/SrTiO$_3$ superlattices have revealed thickness-dependent interface reconstructions arising from polar instability~\cite{yang.ortiz.23}.

\begin{figure*}[]
    \centering
    \includegraphics[width=\linewidth]{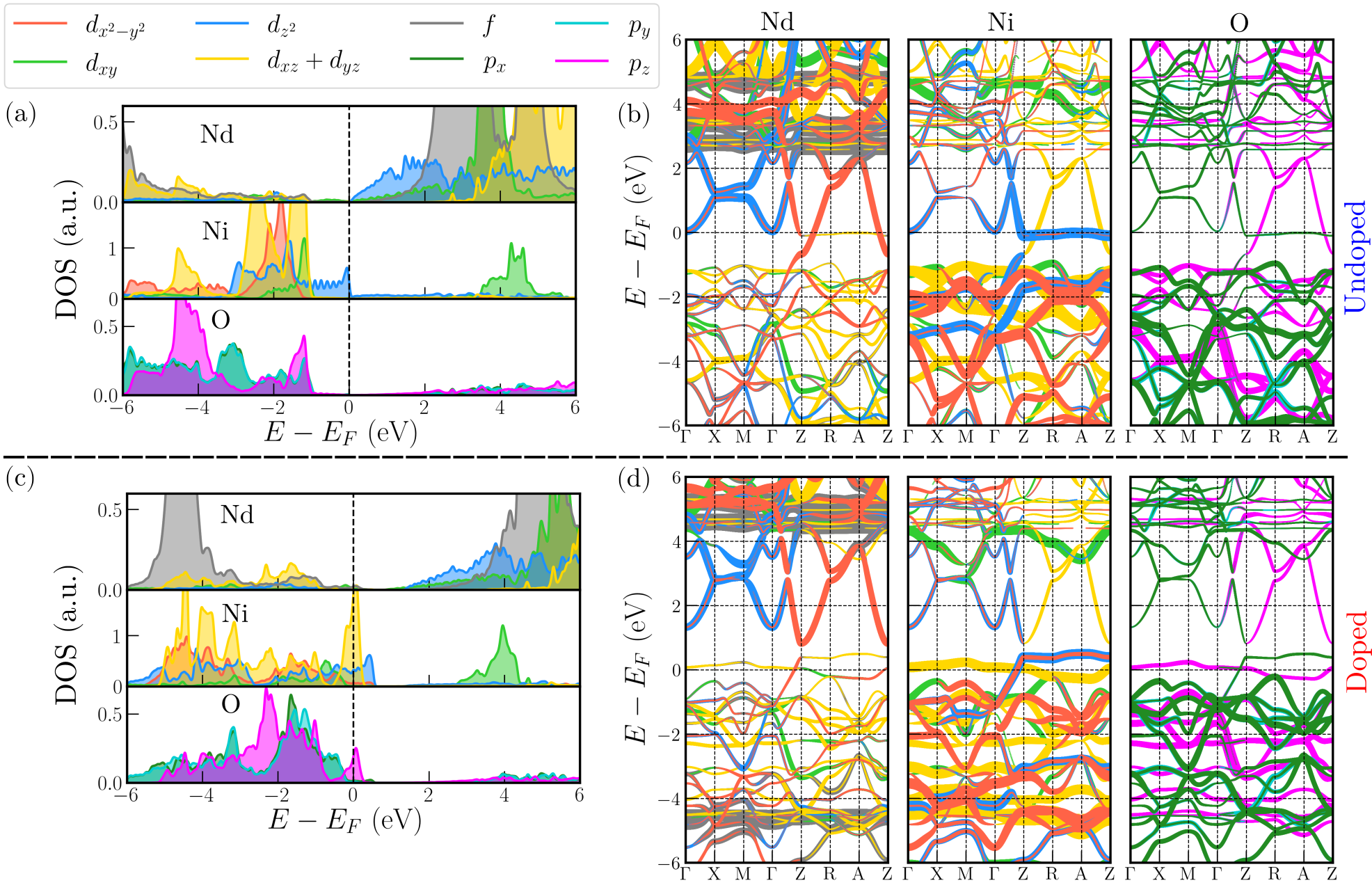} 
    \caption{The atom-projected (a,c) density of states (DOS) and (b,d) band structure of bulk $\mathrm{NdNiO_2}$ with the C-AFM order is shown for two cases: undoped (top), and 20\% hole-doped (bottom).}
    \label{fig.orbital-projected_bands1}
\end{figure*}

In this work, we investigate the electronic and magnetic properties of NdNiO$_2$ within the DFT+$U$ framework, accounting for electron–electron interactions in the Ni $3d$ and Nd $5f$ states. We examine the evolution of the electronic band structure in both bulk systems and thin films of varying thickness. The results for undoped and doped configurations are systematically compared and discussed. Additionally, we assess the impact of SrTiO$_3$
capping layers on the electronic structure and magnetic moments.

The paper is organized as follows. We start by explaining the numerical details in Sec.~\ref{sec.det}. We present and discuss the results on bulk crystals in Sec.~\ref{sec.res.ele}. We then focus on thin films, and discuss electronic and magnetic properties
of the capped an uncapped heterostructures in Sec.~\ref{sec.res.film}. Finally, we summarize the main results in Sec.~\ref{sec.sum}.

\begin{figure*}[]
\centering
\includegraphics[width=0.9\linewidth]{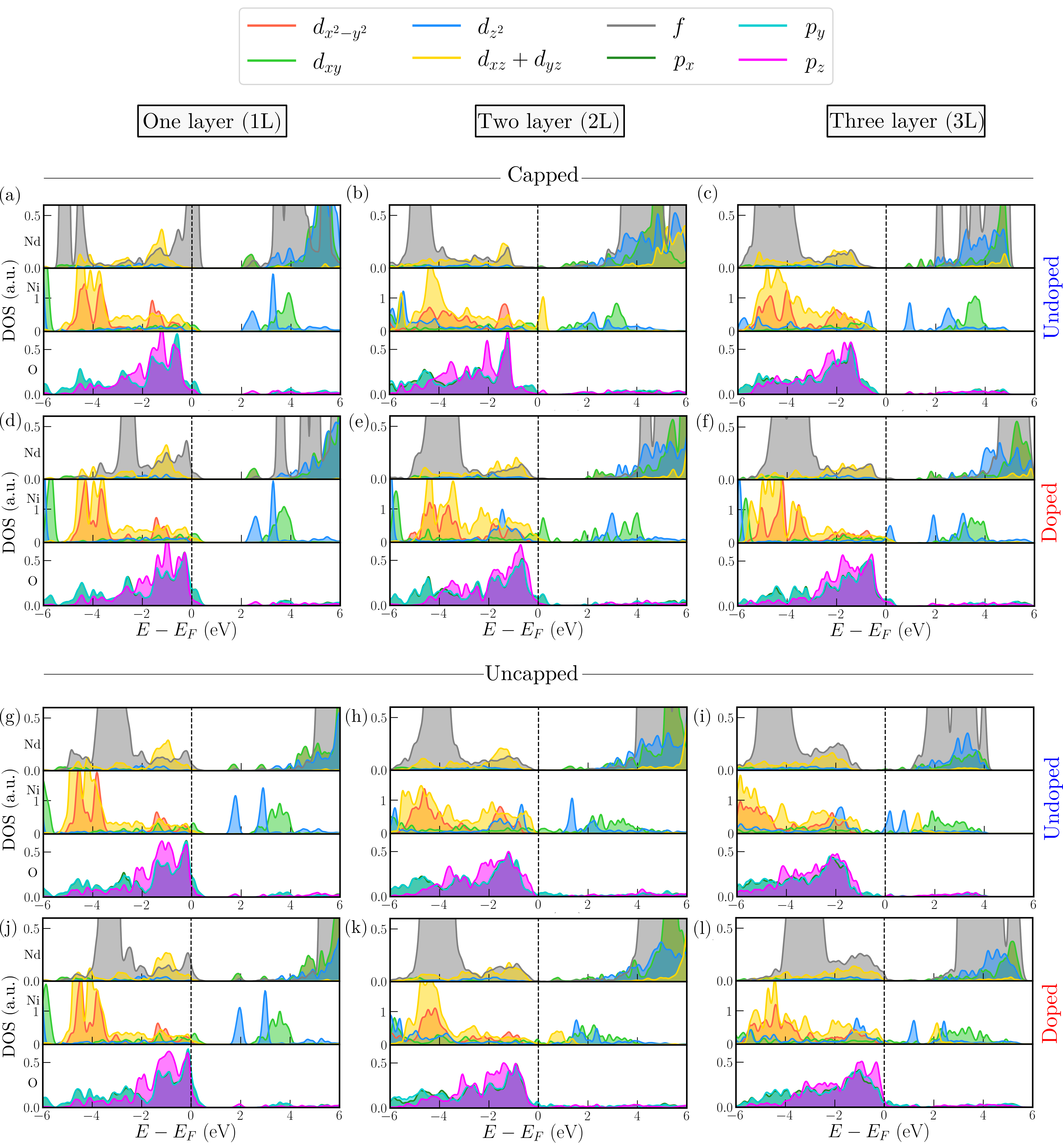}  
\caption{The orbital-resolved density of states of the (a-f) capped and (g-l) uncapped NdNiO$_2$/SrTiO$_3$(001) thin films is shown. The undoped and 20\% hole-doped cases are marked with blue and red texts, respectively.
\label{fig.thin.film}
}
\end{figure*}

\section{Calculation details}
\label{sec.det}

The DFT calculations were performed using the projector augmented-wave (PAW) 
potentials~\cite{blochl.94} implemented in the {\it Vienna Ab initio Simulation Package} 
({\sc Vasp}) code~\cite{kresse.hafner.94,kresse.furthmuller.96,kresse.joubert.99}. 
The exchange-correlation energy was evaluated within the generalized gradient 
approximation (GGA), using the Perdew, Burke, and Ernzerhof (PBE) 
parameterization~\cite{pardew.burke.96}. 
The valence electron configurations for the elements in NdNiO$_{2}$ are as follows: $2s^2p^{4}$ for O, $3d^{9}4s^{1}$ for Ni, and $5s^{2}p^{6}d^{1}6s^{2}$ or $4f^{3.5}5s^{2}p^{6}d^{0.5}6s^{2}$ for Nd with the $f$ orbitals treated as core and valence states, respectively.
The energy cutoff was set to $600$~eV. The local Coulomb interactions for the 
Ni($3d$), Ti($3d$), and Nd($4f$) states were included within the DFT+$U$ 
method~\cite{liechtenstein.95}, using the following parameters, 
Ni: $U=6$~eV and $J=0.7$~eV, Ti: $U=6$~eV and $J=0.5$~eV, 
Nd: $U=8$~eV and $J=1$~eV. 
Similar parameters were used in the earlier studies~\cite{liu.ren.20,wan.ivanov.21,been.lee.21}.

In our investigation, we study three systems presented in Fig.~\ref{fig.cell}: 
the NdNiO$_2$ bulk crystal [Fig.~\ref{fig.cell}(a)], as well as the capped
and uncapped NdNiO$_2$/SrTiO$_3$(001) thin films. The capped 
case is modeled by placing NdNiO$_2$ in between two (bottom and top) 
SrTiO$_3$ layers [see Fig.~\ref{fig.cell}(b)]. The uncapped thin films are studied in the 
supercell with the vacuum layer as shown in Fig.~\ref{fig.cell}(c). 
In all cases, we assume the C-type antiferromagnetic (C-AFM) state
as the starting configuration with the ($\pi,\pi$) spin ordering in the NiO$_2$ plane, 
which shows the lowest energy in the meta-GGA~\cite{zhang.lane.21} and DFT+$U$~\cite{kapeghian.botana.20,ptok.basak.23} studies. 
The C-AFM unit cell is related to the $\sqrt{2}\times\sqrt{2}\times 1$ supercell containing 
two formula units. 
The hole-doping is realized by decreasing the number of electrons by the amount that corresponds to replacing of 20\% of Nd atoms by Sr atoms.
In all of the cases, the {\bf k}-grid in the Monkhorst--Pack scheme~\cite{monkhorst.pack.76} was used. 
As the convergence condition of an optimization loop, we take the energy differences of 
$10^{-5}$~eV and $10^{-7}$~eV for ionic and electronic degrees of freedom, respectively.

\section{Results and discussion}
\label{sec.res}

\subsection{Electronic properties of bulk NdNiO$_2$}
\label{sec.res.ele}

We start our investigation by analyzing the electronic structure of the NdNiO$_2$ bulk crystal 
obtained for the C-AFM ground state. In Fig.~\ref{fig.orbital-projected_bands1}, we present the density of states (DOS) and band structure projected into the Nd($4f$), Nd($5d$), Ni($3d$), and O($2p$) orbitals. Here, the upper and lower panels refer to the undoped and doped case, respectively. In the undoped scenario, a flat band [cf.~Fig.~\ref{fig.orbital-projected_bands1}(b)], mainly originating from Ni($d_{z^2}$) orbital, sits at the Fermi level ($E_F$) along the Z--R--A--Z path. A few states around the $E_F$ indicates that the parent bulk compound is likely a poor metal. This is in well agreement with the previous DFT+$U$~\cite{kapeghian.botana.20,liu.ren.20,wan.ivanov.21,foyevtsova.elfimov.22} and meta-GGA~\cite{zhang.lane.21} studies. Our results with finite $U$ shows considerable difference with the $U=0$ case. For example, the Ni($d_{x^2-y^2}$) band~\cite{sakakibara.usui.20,liu.ren.20,gu.zhu.20,wan.ivanov.21} that crosses $E_F$ for $U=0$, is now located $2$ eV below $E_F$. The other band in the uncorrelated case has Ni($d_{z^2}$) and Nd($d_{z^2}$) character~\cite{liu.ren.20}, which shifts to higher energies in the present DFT+$U$ calculation. We notice several Ni($3d$) states are hybridizing with O$(2p)$ orbital $1$~eV below $E_F$. Finally, we like to mention about the Nd($4f$) states. The large Hubbard interaction of $4f$ orbitals pushes the bands away from $E_F$ compared to the $U=0$ case~\cite{zhang.lane.21,ptok.basak.23}, while the highly localized nature can be seen by the flatness of the bands. 

When the system is $20\%$ hole-doped, the Fermi level shifts to lower energies, modifying the occupations of Ni, Nd, and O bands. We find that the hole doping mainly affects the Ni($d_{z^2}$) states in agreement with the previous DFT+$U$~\cite{wan.ivanov.21,lang.jiang.21} and DFT+DMFT~\cite{lechermann.20b,lechermann.21} studies. Changing of $E_F$ shifts the Ni($d_{yz}+d_{xz}$) orbitals, which creates a new flat band at the Fermi level. There is also a significant shift of oxygen states, especially the $p_z$ orbitals, which are now located at $E_F$. On the other hand, the Nd($5d$) and Nd($4f$) bands are shifted to higher energies above $E_F$. In general, the hole doping increase the occupation of Ni and O bands located close to the Fermi level, which potentially impacts the transport properties of $\mathrm{Nd_{1-x}Sr_{x}NiO_2}$. 

\begin{figure*}[!t]
    \centering
    \includegraphics[width=0.75\linewidth]{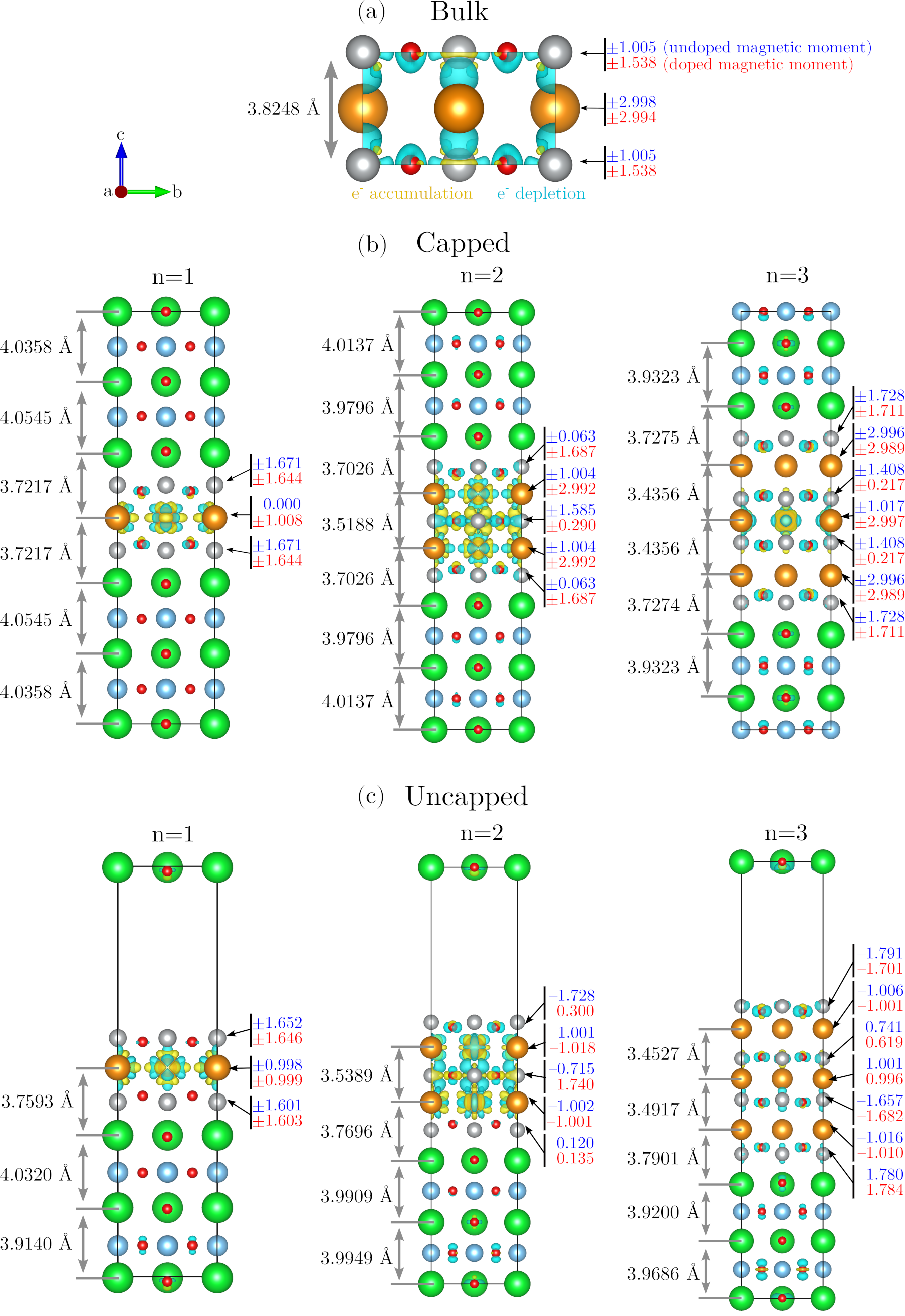} 
    \caption{The figure illustrates charge density ($\Delta \rho$) and magnetic moment differences upon doping for (a) bulk $\mathrm{NdNiO_2}$, (b) capped, and (c) uncapped NdNiO$_2$/SrTiO$_3$(001) thin films. Here, the yellow and cyan regions indicate electron accumulation and depletion, respectively. The distance between atoms along $c$ axis is shown after relaxation. The blue and red number shows magnetic moment (in $\mu_{B}$) for undoped and doped cases, respectively. In all plots, we set the isosurface level to  $0.005~e/\text{\AA}^3$.}
    \label{fig.chg_diff}
\end{figure*}

\subsection{Capped and uncapped NdNiO$_2$/SrTiO$_3$ thin films}
\label{sec.res.film}

In this section, we investigate how capping the thin film and its thickness affect the electronic structure of doped and undoped system. 
In Fig.~\ref{fig.thin.film}, we compare the orbital--resolved DOSs of the thin films with one (1L), two (2L), and three (3L) layers. 

We start by discussing the electronic structure of the undoped systems. For capped 1L, we observe a large contribution of Nd($4f$) states near $E_F$ [cf.~Fig.~\ref{fig.thin.film}(a)], in contrast to the bulk, thicker thin films, or even uncapped 1L film.

In the case of the capped 2L film, the active states are Ni($d_{xz}+d_{yz}$), which hybridize with the O($2p$) states. The electron occupation become much smaller at $E_F$ for the uncapped 2L film. 

Finally, in the 3L case, the Ni and O states shift in energy. Compared to the metallic systems before, the 3L capped film exhibits an energy gap of $\sim 1$ eV at the Fermi level. Such gap does not exist in the uncapped system, where the narrow band of a flat band of Ni ($d_{z^2}$) states is shifted
to the Fermi level. This is comparable to the electronic structure of the bulk crystal, where
($3d_{z^2}$) states are dominant at $E_F$ [see Fig.~\ref{fig.orbital-projected_bands1}(a,b)]. 

In the doped case, the changes are mainly due to the occupation of different Ni($3d$) states. Starting from capped 1L film in Fig.~\ref{fig.thin.film}(d), we find that hole doping significantly reduces the Nd ($4f$) orbitals near $E_F$. Here, the states near $E_F$ mainly consist of $d_{xy}$ orbitals, with a small contribution from $d_{xz+yz}$ orbitals. Such Ni ($3d$) states hybridize with the O ($2p$) states. The electronic landscape are similar for the uncapped film.

As we move to 2L capped system in  Fig.~\ref{fig.thin.film}(e), the Ni $3d_{xy}$ contribution rises while $3d_{xz+yz}$ states get suppressed. The enhancement of $3d_{xy}$ is only seen in the capped film, and is not present in the uncapped version shown in Fig.~\ref{fig.thin.film}(k). 

Finally, we describe the 3L system, which is closer to the bulk crystal. In the capped scenario, the parent compound is an insulator, which under hole doping [cf.~Fig.~\ref{fig.thin.film}(f)] becomes metallic since Ni ($3d$) and O ($2p$) states move near the Fermi level. We find that the states near $E_F$ are mainly occupied by the $3d_{z^2}$ and $3d_{yz+xz}$ orbitals -- a feature similar to the bulk results in Fig.~\ref{fig.orbital-projected_bands1}, where $3d_{z^2}$ state lead to a fat flat band near $E_F$. Such correspondence with the bulk crystal is not seen in the uncapped counterpart, and thus, the result is more non-trivial. Under hole-doping, the $3d_{z^2}$ peak [see Fig.~\ref{fig.thin.film}(l)] of the uncapped film is shifted to higher energies, and the small number of states near $E_F$ has contributions from $d_{xz+yz}$ and $d_{z^2}$ orbitals. Furthermore, the hole doping increased the Ni-O hybridization extent.

In Fig.~\ref{fig.chg_diff}, we present the changes of magnetic moments and charge density differences induced by hole doping for both the bulk and thin films. In the bulk, the magnetic moments on Ni atoms are largely enhanced due to doping by about 50$\%$, while on Nd atoms they are modified very weakly. The changes in electron charge density are found only in the Ni-O planes. Charge decreasing induced by the hole doping is related mainly with the $z$ component of the orbitals, i.e. the Ni($d_{z^2}$) and O($p_z$) orbitals, which results from the changes in the electronic structure presented in Fig.~\ref{fig.orbital-projected_bands1}.  
There is also some charge depletion in the bonds between Ni and O atoms.   

In contrast to changes found in the bulk crystal, the main differences between the undoped and doped 1L system is found close to the Nd atoms. For the capped films, we observe the redistribution of charge density around the Nd and O atoms, while there are no changes found at the Ni atoms. Interestingly, in the uncapped case, the charge density is modified mostly close to the Nd Atoms.  
In the capped system, the magnetic moments on Nd is equal zero for the undoped case, while after doping it is close to $\pm 1$~$\mu_\text{B}$. 
In the uncapped thin film, the Nd magnetic moments are almost the same with and without doping.
The magnetic moments on the Ni atoms have very similar values in the capped and uncapped films ($\sim 1.6$~$\mu_\text{B}$), and they are only slightly reduced due to doping.

In the 2L systems, the charge distribution around the Nd atoms looks similar as in the 1L films, however, the magnetic moments are strongly enhanced, reaching values $\sim 3$~$\mu_\text{B}$. In two Ni-O planes bordering on SrTiO$_3$, the charge distribution is also very similar to the 1L case, but it is strongly modified in the middle plane. A significant charge modification is found in the $d_{xz}$, $d_{yz}$, and $d_{xy}$ orbitals, but we observe an opposite behavior in the capped and uncapped films. In the capped film, there is charge accumulation in the $d_{xz}+d_{yz}$ orbitals and charge depletion in the $d_{xy}$ orbitals.
This opposite behavior can be explained by analyzing the shifts of the bands induced by doping in the electronic density of states 
presented in Fig.~\ref{fig.thin.film}. There is also a significant difference in the arrangement of magnetic moments in these two systems.
In the capped film, there is the AFM order in each Ni-O plane, but the values of magnetic moments are strongly modified by doping.
In the middle plane, the magnetic moments on the Ni atoms are strongly reduced, while in in the two other planes we observe a change from a very small
value to about 1.7~$\mu_\text{B}$. Interestingly, the magnetic order differs in the uncapped system. Here, we observe the FM configurations in each Ni-O plane,
but the magnetic moments are opposite in the adjacent layers. We can also notice that the magnetic moments are reversed due to doping in two planes lying closer to the vacuum layer. In these two planes, the changes in the magnitudes of magnetic moments are the strongest.
In the plane bordering on SrTiO$_3$, the magnetic moments are almost no altered and amount to only $\sim$0.1~$\mu_\text{B}$.
Due to such modifications, the total magnetization of the film is not equal zero, and its value depends on doping.   

We observe different changes of the charge density in the 3L systems. The modifications around the Nd atoms are much weaker than in the thinner layers.
Only in the middle layer there is some charge redistribution around Nd, which is connect with the large increase of the magnetic moment.
In the two other layers, there is no change in charge density at the Nd sites, exactly as in the bulk.
The changes around the Ni atoms are much weaker than in the bulk, and they depend on the position of the layer.
In the capped film, we observe charge depletion in the Ni($d_{z^2}$) orbitals in two middle planes, while in the planes bordering on SrTiO$_3$ no charge modification is observed. Concerning the oxygen atoms, there is electron density decrease in the $p_x$ and $p_y$ orbitals in all planes, and no charge modification in the $p_z$ orbitals. 
In the uncapped system, the changes are similar but the weakest affect is found in the surface layer. 
Similarly to the magnetic orders in the 2L films, we observe large differences between the capped and uncapped systems.
In the capped film, the magnetic moments on the Ni atoms are strongly reduced due to doping in two middle Ni-O planes, while they are only weakly modified in two other planes. Like in the 2L systems, we observe the FM order within the planes with the opposite magnetic moments in the neigboring planes.
However, in contrast to the 2L system, the magnetic moments are not reversed by doping. 

In all thin films, apart from the changes in the electronic properties, we observe also the structural reconstruction, which induces
the changes of atomic positions comparing to the geometry of the bulk crystal.
The distances between the layers are strongly reduced mainly in the middle parts of the films in the capped systems and close to the surface
in the uncapped ones.
Additionally, there are shifts of oxygen atoms along the $z$ direction, which depend on the position of the layer.
In the outer Ni-O planes, the atoms are shifted towards the center of the film,
while in the middle planes of the 3L systems, these shifts are in opposite direction and with smaller amplitudes.
These reconstructions can be connected with other changes in the electronic and atomic structure of the NdNiO$_2$/SrTiO$_3$ interface induced by the polar instability~\cite{geisler.pentcheva.20,he.jiang.20,zhang.lin.20,goodge.geisler.23,yang.ortiz.23}.

\section{Summary}
\label{sec.sum}

Using the DFT+$U$ approach, we studied the electronic and magnetic properties of the bulk crystal and thin films of NdNiO$_2$.
The effect of hole doping was studied by decresing the number of electrons in the studied systems. 
In the bulk, the main changes induced by doping are connected with the Ni($d_{z^2}$) orbitals, which dominate at the Fermi energy.
Additionally, we observe the activation of the $d_{xz}$ and $d_{yz}$ orbitals.
Due to the large Coulomb interaction $U$, the Nd($4f$) states are located about 2~eV above $E_\text{F}$.
In the undoped system, the Nd($d_{z^2}$) band is lying just above the Fermi energy, and it is shifted to higher energies with doping.
The magnetic moments on the Ni atoms are increased by about 50\% due to doping, while on the Nd atoms they are not modified.   

In the thin films, we observe significant changes in the electronic structure and magnetic properties, which depend on the number of layers
and the capping. Comparing to the bulk, we observe the shift of the Ni($d_{xy}$, $d_{xz}$, $d_{yz}$) and O($p_x$, $p_y$, $p_z$) states close to 
the Fermi energy. The position of the Ni($d_{z^2}$) states, which play important role in the bulk material, 
strongly depend on the number of layers and they are located close to $E_\text{F}$ in the thicker (3L) films.
The Nd($4f$) orbitals are found at the Fermi energy only in the 1L capped film, while in the other systems they are shifted to higher energies by about 2~eV. Also, the Nd($d_{z^2}$) states are located higher in energy comparing to the bulk crystal. 
We revealed also differences in the charge distribution and magnetic moments when comparing the results for the bulk and film films.
In the 1L and 2L systems, we observe strong charge redistribution and the increase of magnetic moments on the Nd atoms.
The effect of doping in the Ni-O planes depends on the number of layers and capping.
In the uncapped 2L and 3L systems, we found the change of magnetic order on the Ni atoms, which leads
to a non-zero value of the total magnetization, in contrast to the bulk and capped films. 
The observed electronic and structural reconstructions modify the polar properties of thin films, thus influencing
stability conditions of the interface.

\begin{acknowledgments}
Some figures in this work were rendered using {\sc Vesta}~\cite{momma.izumi.11} software.
It is a pleasure to thank A. Greco for insightful discussion.
We kindly acknowledge support from the National Science Centre 
(NCN, Poland) under Project No.~2021/43/B/ST3/02166. 
\end{acknowledgments}

\bibliography{biblio}

@article{abadi.xu.25,
  title = {Electronic Structure of the Alternating Monolayer-Trilayer Phase of ${\mathrm{La}}_{3}{\text{Ni}}_{2}{\mathrm{O}}_{7}$},
  author = {Abadi, Sebastien and Xu, Ke-Jun and Lomeli, Eder G. and Puphal, Pascal and Isobe, Masahiko and Zhong, Yong and Fedorov, Alexei V. and Mo, Sung-Kwan and Hashimoto, Makoto and Lu, Dong-Hui and Moritz, Brian and Keimer, Bernhard and Devereaux, Thomas P. and Hepting, Matthias and Shen, Zhi-Xun},
  journal = {Phys. Rev. Lett.},
  volume = {134},
  issue = {12},
  pages = {126001},
  numpages = {7},
  year = {2025},
  month = {Mar},
  publisher = {American Physical Society},
  doi = {10.1103/PhysRevLett.134.126001},
  url = {https://link.aps.org/doi/10.1103/PhysRevLett.134.126001}
}

@article{Adh20,
  title = {Orbital-selective superconductivity in a two-band model of infinite-layer nickelates},
  author = {Adhikary, Priyo and Bandyopadhyay, Subhadeep and Das, Tanmoy and Dasgupta, Indra and Saha-Dasgupta, Tanusri},
  journal = {Phys. Rev. B},
  volume = {102},
  pages = {100501},
  year = {2020},
  publisher = {American Physical Society},
  doi = {10.1103/PhysRevB.102.100501},
  url = {https://doi.org/10.1103/PhysRevB.102.100501}
}

@article{been.lee.21,
  title = {Electronic Structure Trends Across the Rare-Earth Series in Superconducting Infinite-Layer Nickelates},
  author = {Been, Emily and Lee, Wei-Sheng and Hwang, Harold Y. and Cui, Yi and Zaanen, Jan and Devereaux, Thomas and Moritz, Brian and Jia, Chunjing},
  journal = {Phys. Rev. X},
  volume = {11},
  issue = {1},
  pages = {011050},
  numpages = {17},
  year = {2021},
  month = {Mar},
  publisher = {American Physical Society},
  doi = {10.1103/PhysRevX.11.011050},
  url = {https://doi.org/10.1103/PhysRevX.11.011050}
}

@article{berit.danfend.21,
author = {Berit H. Goodge  and Danfeng Li  and Kyuho Lee  and Motoki Osada  and Bai Yang Wang  and George A. Sawatzky  and Harold Y. Hwang  and Lena F. Kourkoutis },
title = {Doping evolution of the {Mott--Hubbard} landscape in infinite-layer nickelates},
journal = {PNAS},
volume = {118},
number = {2},
pages = {e2007683118},
year = {2021},
doi = {10.1073/pnas.2007683118},
URL = {https://www.pnas.org/doi/abs/10.1073/pnas.2007683118}
}

@article{blochl.94,
  title = {Projector augmented-wave method},
  author = {Bl\"ochl, P. E.},
  journal = {Phys. Rev. B},
  volume = {50},
  issue = {24},
  pages = {17953},
  numpages = {0},
  year = {1994},
  month = {Dec},
  publisher = {American Physical Society},
  doi = {10.1103/PhysRevB.50.17953},
  url = {http://doi.org/10.1103/PhysRevB.50.17953}
}

@article{botana.norman.20,
  title = {Similarities and Differences between {LaNiO$_{2}$} and {CaCuO$_{2}$} and Implications for Superconductivity},
  author = {Botana, A. S. and Norman, M. R.},
  journal = {Phys. Rev. X},
  volume = {10},
  issue = {1},
  pages = {011024},
  numpages = {6},
  year = {2020},
  month = {Feb},
  publisher = {American Physical Society},
  doi = {10.1103/PhysRevX.10.011024},
  url = {https://doi.org/10.1103/PhysRevX.10.011024}
}

@article{chen.jiang.22,
  title = {Magnetism in doped infinite-layer {NdNiO$_{2}$} studied by combined density functional theory and dynamical mean-field theory},
  author = {Chen, Dachuan and Jiang, Peiheng and Si, Liang and Lu, Yi and Zhong, Zhicheng},
  journal = {Phys. Rev. B},
  volume = {106},
  issue = {4},
  pages = {045105},
  numpages = {11},
  year = {2022},
  month = {Jul},
  publisher = {American Physical Society},
  doi = {10.1103/PhysRevB.106.045105},
  url = {https://doi.org/10.1103/PhysRevB.106.045105}
}

@article{Cho20,
  title = {Role of $4f$ states in infinite-layer {NdNiO$_{2}$}},
  author = {Choi, Mi-Young and Lee, Kwan-Woo and Pickett, Warren E.},
  journal = {Phys. Rev. B},
  volume = {101},
  pages = {020503},
  year = {2020},
  publisher = {American Physical Society},
  doi = {10.1103/PhysRevB.101.020503},
  url = {https://doi.org/10.1103/PhysRevB.101.020503}
}

@article{deng.jiang.21,
doi = {10.1209/0295-5075/ac2073},
url = {https://dx.doi.org/10.1209/0295-5075/ac2073},
year = {2021},
month = {nov},
publisher = {EDP Sciences, IOP Publishing and Società Italiana di Fisica},
volume = {135},
number = {6},
pages = {67001},
author = {Fenglin Deng and Peiheng Jiang and Yi Lu and Zhicheng Zhong},
title = {First-principle study of {Sr}-doping effect in {Nd$_{1-x}$Sr$_{x}$NiO$_{2}$}},
journal = {EPL}
}

@article{Pli22,
  title = {Screening in a two-band model for superconducting infinite-layer nickelate},
  author = {Plienbumrung, Tharathep and Daghofer, Maria and Schmid, Michael and Ole\'{s}, Andrzej M.},
  journal = {Phys. Rev. B},
  volume = {106},
  pages = {134504},
  numpages = {10},
  year = {2022},
  publisher = {American Physical Society},
  doi = {10.1103/PhysRevB.106.134504},
  url = {https://doi.org/10.1103/PhysRevB.106.134504}
}

@Article{fowlie.hadjimichael.22,
author={Fowlie, Jennifer and Hadjimichael, Marios and Martins, Maria M.
and Li, Danfeng and Osada, Motoki and Wang, Bai Yang and Lee, Kyuho
and Lee, Yonghun and Salman, Zaher and Prokscha, Thomas and Triscone, Jean-Marc
and Hwang, Harold Y. and Suter, Andreas},
title={Intrinsic magnetism in superconducting infinite-layer nickelates},
journal={Nat. Phys.},
year={2022},
month={Sep},
day={01},
volume={18},
number={9},
pages={1043-1047},
issn={1745-2481},
doi={10.1038/s41567-022-01684-y},
url={https://doi.org/10.1038/s41567-022-01684-y}
}

@article{foyevtsova.elfimov.22,
  title = {Distinct electridelike nature of infinite-layer nickelates and the resulting theoretical challenges to calculate their electronic structure},
  author = {Foyevtsova, Kateryna and Elfimov, Ilya and Sawatzky, George A.},
  journal = {Phys. Rev. B},
  volume = {108},
  issue = {20},
  pages = {205124},
  numpages = {11},
  year = {2023},
  month = {Nov},
  publisher = {American Physical Society},
  doi = {10.1103/PhysRevB.108.205124},
  url = {https://doi.org/10.1103/PhysRevB.108.205124}
}

@article{geisler.pentcheva.20,
  title = {Fundamental difference in the electronic reconstruction of infinite-layer versus perovskite neodymium nickelate films on {SrTiO$_{3}$(001)}},
  author = {Geisler, Benjamin and Pentcheva, Rossitza},
  journal = {Phys. Rev. B},
  volume = {102},
  issue = {2},
  pages = {020502},
  numpages = {6},
  year = {2020},
  month = {Jul},
  publisher = {American Physical Society},
  doi = {10.1103/PhysRevB.102.020502},
  url = {https://doi.org/10.1103/PhysRevB.102.020502}
}

@Article{goodge.geisler.23,
author={Goodge, Berit H.
and Geisler, Benjamin
and Lee, Kyuho
and Osada, Motoki
and Wang, Bai Yang
and Li, Danfeng
and Hwang, Harold Y.
and Pentcheva, Rossitza
and Kourkoutis, Lena F.},
title={Resolving the polar interface of infinite-layer nickelate thin films},
journal={Nature Materials},
year={2023},
month={Apr},
day={01},
volume={22},
number={4},
pages={466-473},
issn={1476-4660},
doi={10.1038/s41563-023-01510-7},
url={https://doi.org/10.1038/s41563-023-01510-7}
}

@Article{gu.li.20,
author={Gu, Qiangqiang and Li, Yueying and Wan, Siyuan
and Li, Huazhou and Guo, Wei and Yang, Huan
and Li, Qing and Zhu, Xiyu and Pan, Xiaoqing
and Nie, Yuefeng and Wen, Hai-Hu}, 
title={Single particle tunneling spectrum of superconducting {Nd$_{1-x}$Sr$_{x}$NiO$_{2}$} thin films},
journal={Nature Commun.},
year={2020},
month={Nov},
day={27},
volume={11},
number={1},
pages={6027},
issn={2041-1723},
doi={10.1038/s41467-020-19908-1},
url={https://doi.org/10.1038/s41467-020-19908-1}
}

@Article{gu.zhu.20,
author={Gu, Yuhao
and Zhu, Sichen
and Wang, Xiaoxuan
and Hu, Jiangping
and Chen, Hanghui},
title={A substantial hybridization between correlated {Ni}-$d$ orbital and itinerant electrons in infinite-layer nickelates},
journal={Commun. Phys.},
year={2020},
month={May},
day={14},
volume={3},
number={1},
pages={84},
issn={2399-3650},
doi={10.1038/s42005-020-0347-x},
url={https://doi.org/10.1038/s42005-020-0347-x}
}

@article{he.jiang.20,
  title = {Polarity-induced electronic and atomic reconstruction at {NdNiO$_{2}$/SrTiO$_{3}$} interfaces},
  author = {He, Ri and Jiang, Peiheng and Lu, Yi and Song, Yidao and Chen, Mingxing and Jin, Mingliang and Shui, Lingling and Zhong, Zhicheng},
  journal = {Phys. Rev. B},
  volume = {102},
  issue = {3},
  pages = {035118},
  numpages = {7},
  year = {2020},
  month = {Jul},
  publisher = {American Physical Society},
  doi = {10.1103/PhysRevB.102.035118},
  url = {https://doi.org/10.1103/PhysRevB.102.035118}
}

@misc{huang.li.25,
      title={Superconductivity in monolayer-trilayer phase of La$_3$Ni$_2$O$_7$ under high pressure}, 
      author={Chaoxin Huang and Jingyuan Li and Xing Huang and Hengyuan Zhang and Deyuan Hu and Mengwu Huo and Xiang Chen and Zhen Chen and Hualei Sun and Meng Wang},
      year={2025},
      eprint={2510.12250},
      archivePrefix={arXiv},
      primaryClass={cond-mat.supr-con},
      url={https://arxiv.org/abs/2510.12250}, 
}

@article{jiang.si.19,
  title = {Electronic structure of rare-earth infinite-layer {$R$NiO$_{2}$} ({$R=$La, Nd})},
  author = {Jiang, Peiheng and Si, Liang and Liao, Zhaoliang and Zhong, Zhicheng},
  journal = {Phys. Rev. B},
  volume = {100},
  issue = {20},
  pages = {201106},
  numpages = {5},
  year = {2019},
  month = {Nov},
  publisher = {American Physical Society},
  doi = {10.1103/PhysRevB.100.201106},
  url = {https://doi.org/10.1103/PhysRevB.100.201106}
}

@article{kapeghian.botana.20,
  title = {Electronic structure and magnetism in infinite-layer nickelates {$R$NiO$_{2}$} ({$R=$La-Lu})},
  author = {Kapeghian, Jesse and Botana, Antia S.},
  journal = {Phys. Rev. B},
  volume = {102},
  issue = {20},
  pages = {205130},
  numpages = {14},
  year = {2020},
  month = {Nov},
  publisher = {American Physical Society},
  doi = {10.1103/PhysRevB.102.205130},
  url = {https://doi.org/10.1103/PhysRevB.102.205130}
}

@article{karp.botana.20,
  title = {Many-Body Electronic Structure of {NdNiO$_{2}$} and {CaCuO$_{2}$}},
  author = {Karp, Jonathan and Botana, Antia S. and Norman, Michael R. and Park, Hyowon and Zingl, Manuel and Millis, Andrew},
  journal = {Phys. Rev. X},
  volume = {10},
  issue = {2},
  pages = {021061},
  numpages = {11},
  year = {2020},
  month = {Jun},
  publisher = {American Physical Society},
  doi = {10.1103/PhysRevX.10.021061},
  url = {https://doi.org/10.1103/PhysRevX.10.021061}
}

@article{karp.hampel.22,
  title = {Superconductivity and antiferromagnetism in {NdNiO$_{2}$} and {CaCuO$_{2}$}: A cluster {DMFT} study},
  author = {Karp, Jonathan and Hampel, Alexander and Millis, Andrew J.},
  journal = {Phys. Rev. B},
  volume = {105},
  issue = {20},
  pages = {205131},
  numpages = {11},
  year = {2022},
  month = {May},
  publisher = {American Physical Society},
  doi = {10.1103/PhysRevB.105.205131},
  url = {https://doi.org/10.1103/PhysRevB.105.205131}
}

@Article{kitatani.si.20,
author={Kitatani, Motoharu
and Si, Liang
and Janson, Oleg
and Arita, Ryotaro
and Zhong, Zhicheng
and Held, Karsten},
title={Nickelate superconductors--a renaissance of the one-band {Hubbard} model},
journal={npj Quantum Mater.},
year={2020},
month={Aug},
day={21},
volume={5},
number={1},
pages={59},
issn={2397-4648},
doi={10.1038/s41535-020-00260-y},
url={https://doi.org/10.1038/s41535-020-00260-y}
}

@article{kresse.furthmuller.96,
  title = {Efficient iterative schemes for ab initio total-energy calculations using a plane-wave basis set},
  author = {Kresse, G. and Furthm\"uller, J.},
  journal = {Phys. Rev. B},
  volume = {54},
  issue = {16},
  pages = {11169},
  numpages = {0},
  year = {1996},
  month = {Oct},
  publisher = {American Physical Society},
  doi = {10.1103/PhysRevB.54.11169},
  url = {http://doi.org/10.1103/PhysRevB.54.11169}
}

@article{kresse.hafner.94,
  title = {Ab initio molecular-dynamics simulation of the liquid-metal--amorphous-semiconductor transition in germanium},
  author = {Kresse, G. and Hafner, J.},
  journal = {Phys. Rev. B},
  volume = {49},
  issue = {20},
  pages = {14251},
  numpages = {0},
  year = {1994},
  month = {May},
  publisher = {American Physical Society},
  doi = {10.1103/PhysRevB.49.14251},
  url = {http://doi.org/10.1103/PhysRevB.49.14251}
}

@article{kresse.joubert.99,
  title = {From ultrasoft pseudopotentials to the projector augmented-wave method},
  author = {Kresse, G. and Joubert, D.},
  journal = {Phys. Rev. B},
  volume = {59},
  issue = {3},
  pages = {1758},
  numpages = {0},
  year = {1999},
  month = {Jan},
  publisher = {American Physical Society},
  doi = {10.1103/PhysRevB.59.1758},
  url = {http://doi.org/10.1103/PhysRevB.59.1758}
}

@article{krieger.martinelli.22,
  title = {Charge and Spin Order Dichotomy in {NdNiO$_{2}$} Driven by the Capping Layer},
  author = {Krieger, G. and Martinelli, L. and Zeng, S. and Chow, L. E. and Kummer, K. and Arpaia, R. and Moretti Sala, M. and Brookes, N. B. and Ariando, A. and Viart, N. and Salluzzo, M. and Ghiringhelli, G. and Preziosi, D.},
  journal = {Phys. Rev. Lett.},
  volume = {129},
  issue = {2},
  pages = {027002},
  numpages = {7},
  year = {2022},
  month = {Jul},
  publisher = {American Physical Society},
  doi = {10.1103/PhysRevLett.129.027002},
  url = {https://doi.org/10.1103/PhysRevLett.129.027002}
}

@article{lang.jiang.21,
  title = {Strongly correlated doped hole carriers in the superconducting nickelates: Their location, local many-body state, and low-energy effective {Hamiltonian}},
  author = {Lang, Zi-Jian and Jiang, Ruoshi and Ku, Wei},
  journal = {Phys. Rev. B},
  volume = {103},
  issue = {18},
  pages = {L180502},
  numpages = {6},
  year = {2021},
  month = {May},
  publisher = {American Physical Society},
  doi = {10.1103/PhysRevB.103.L180502},
  url = {https://doi.org/10.1103/PhysRevB.103.L180502}
}

@article{lechermann.20,
  title = {Late transition metal oxides with infinite-layer structure: Nickelates versus cuprates},
  author = {Lechermann, Frank},
  journal = {Phys. Rev. B},
  volume = {101},
  issue = {8},
  pages = {081110},
  numpages = {5},
  year = {2020},
  month = {Feb},
  publisher = {American Physical Society},
  doi = {10.1103/PhysRevB.101.081110},
  url = {https://doi.org/10.1103/PhysRevB.101.081110}
}

@article{lechermann.20b,
  title = {Multiorbital Processes Rule the {Nd$_{1-x}$Sr$_{x}$NiO$_{2}$} Normal State},
  author = {Lechermann, Frank},
  journal = {Phys. Rev. X},
  volume = {10},
  issue = {4},
  pages = {041002},
  numpages = {14},
  year = {2020},
  month = {Oct},
  publisher = {American Physical Society},
  doi = {10.1103/PhysRevX.10.041002},
  url = {https://doi.org/10.1103/PhysRevX.10.041002}
}

@article{lechermann.21,
  title = {Doping-dependent character and possible magnetic ordering of {NdNiO$_{2}$}},
  author = {Lechermann, Frank},
  journal = {Phys. Rev. Mater.},
  volume = {5},
  issue = {4},
  pages = {044803},
  numpages = {13},
  year = {2021},
  month = {Apr},
  publisher = {American Physical Society},
  doi = {10.1103/PhysRevMaterials.5.044803},
  url = {https://doi.org/10.1103/PhysRevMaterials.5.044803}
}

@article{leonov.skornyakov.20,
  title = {Lifshitz transition and frustration of magnetic moments in infinite-layer {NdNiO$_{2}$} upon hole doping},
  author = {Leonov, I. and Skornyakov, S. L. and Savrasov, S. Y.},
  journal = {Phys. Rev. B},
  volume = {101},
  issue = {24},
  pages = {241108},
  numpages = {5},
  year = {2020},
  month = {Jun},
  publisher = {American Physical Society},
  doi = {10.1103/PhysRevB.101.241108},
  url = {https://doi.org/10.1103/PhysRevB.101.241108}
}

@Article{liu.ren.20,
author={Liu, Zhao
and Ren, Zhi
and Zhu, Wei
and Wang, Zhengfei
and Yang, Jinlong},
title={Electronic and magnetic structure of infinite-layer {NdNiO$_{2}$}: trace of antiferromagnetic metal},
journal={npj Quantum Mater.},
year={2020},
month={May},
day={15},
volume={5},
number={1},
pages={31},
issn={2397-4648},
doi={10.1038/s41535-020-0229-1},
url={https://doi.org/10.1038/s41535-020-0229-1}
}

@article{liu.xu.21,
  title = {Doping dependence of electronic structure of infinite-layer {NdNiO$_{2}$}},
  author = {Liu, Zhao and Xu, Chenchao and Cao, Chao and Zhu, W. and Wang, Z. F. and Yang, Jinlong},
  journal = {Phys. Rev. B},
  volume = {103},
  issue = {4},
  pages = {045103},
  numpages = {10},
  year = {2021},
  month = {Jan},
  publisher = {American Physical Society},
  doi = {10.1103/PhysRevB.103.045103},
  url = {https://doi.org/10.1103/PhysRevB.103.045103}
}

@Article{li.he.20,
author={Li, Qing
and He, Chengping
and Si, Jin
and Zhu, Xiyu
and Zhang, Yue
and Wen, Hai-Hu},
title={Absence of superconductivity in bulk {Nd$_{1-x}$Sr$_{x}$NiO$_{2}$}},
journal={Commun. Mater.},
year={2020},
month={Apr},
day={03},
volume={1},
number={1},
pages={16},
issn={2662-4443},
doi={10.1038/s43246-020-0018-1},
url={https://doi.org/10.1038/s43246-020-0018-1}
}

@Article{li.lee.19,
author={Li, Danfeng
and Lee, Kyuho
and Wang, Bai Yang
and Osada, Motoki
and Crossley, Samuel
and Lee, Hye Ryoung
and Cui, Yi
and Hikita, Yasuyuki
and Hwang, Harold Y.},
title={Superconductivity in an infinite-layer nickelate},
journal={Nature},
year={2019},
month={Aug},
day={01},
volume={572},
number={7771},
pages={624},
issn={1476-4687},
doi={10.1038/s41586-019-1496-5},
url={https://doi.org/10.1038/s41586-019-1496-5}
}

@article{liechtenstein.95,
  title={Density-functional theory and strong interactions: Orbital ordering in {Mott-Hubbard} insulators},
  author={Liechtenstein, A. I. and Anisimov, V. I. and Zaanen, J.},
  journal = {Phys. Rev. B},
  volume = {52},
  issue = {8},
  pages = {R5467--R5470},
  numpages = {0},
  year = {1995},
  month = {Aug},
  publisher = {American Physical Society},
  doi = {10.1103/PhysRevB.52.R5467},
  url = {https://doi.org/10.1103/PhysRevB.52.R5467}
}

@article{lu.rossi.21,
author = {H. Lu  and M. Rossi  and A. Nag  and M. Osada  and D. F. Li  and K. Lee  and B. Y. Wang  and M. Garcia-Fernandez  and S. Agrestini  and Z. X. Shen  and E. M. Been  and B. Moritz  and T. P. Devereaux  and J. Zaanen  and H. Y. Hwang  and Ke-Jin Zhou and W. S. Lee },
title = {Magnetic excitations in infinite-layer nickelates},
journal = {Science},
volume = {373},
number = {6551},
pages = {213},
year = {2021},
doi = {10.1126/science.abd7726},
URL = {https://www.science.org/doi/abs/10.1126/science.abd7726}
}

@article{momma.izumi.11,
author = "Momma, K. and Izumi, F.",
title = "{{\sc Vesta3} for three-dimensional visualization of crystal, volumetric and morphology data}",
journal = "J. Appl. Crystallogr.",
year = "2011",
volume = "44",
number = "6",
pages = "1272",
month = "Dec",
doi = {10.1107/S0021889811038970},
url = {https://doi.org/10.1107/S0021889811038970}
}

@article{monkhorst.pack.76,
  title = {Special points for {Brillouin}-zone integrations},
  author = {Monkhorst, H. J. and Pack, J. D.},
  journal = {Phys. Rev. B},
  volume = {13},
  issue = {12},
  pages = {5188},
  numpages = {0},
  year = {1976},
  month = {Jun},
  publisher = {American Physical Society},
  doi = {10.1103/PhysRevB.13.5188},
  url = {http://doi.org/10.1103/PhysRevB.13.5188}
}

@article{nomura.arita.22,
doi = {10.1088/1361-6633/ac5a60},
url = {https://dx.doi.org/10.1088/1361-6633/ac5a60},
year = {2022},
month = {mar},
publisher = {IOP Publishing},
volume = {85},
number = {5},
pages = {052501},
author = {Yusuke Nomura and Ryotaro Arita},
title = {Superconductivity in infinite-layer nickelates},
journal = {Rep. Prog. Phys.}
}

@article{nomura.hirayama.19,
  title = {Formation of a two-dimensional single-component correlated electron system and band engineering in the nickelate superconductor {NdNiO$_{2}$}},
  author = {Nomura, Yusuke and Hirayama, Motoaki and Tadano, Terumasa and Yoshimoto, Yoshihide and Nakamura, Kazuma and Arita, Ryotaro},
  journal = {Phys. Rev. B},
  volume = {100},
  issue = {20},
  pages = {205138},
  numpages = {11},
  year = {2019},
  month = {Nov},
  publisher = {American Physical Society},
  doi = {10.1103/PhysRevB.100.205138},
  url = {https://doi.org/10.1103/PhysRevB.100.205138}
}

@article{ortiz.puphal.22,
  title = {Magnetic correlations in infinite-layer nickelates: An experimental and theoretical multimethod study},
  author = {Ortiz, R. A. and Puphal, P. and Klett, M. and Hotz, F. and Kremer, R. K. and Trepka, H. and Hemmida, M. and von Nidda, H.-A. Krug and Isobe, M. and Khasanov, R. and Luetkens, H. and Hansmann, P. and Keimer, B. and Sch\"{a}fer, T. and Hepting, M.},
  journal = {Phys. Rev. Research},
  volume = {4},
  issue = {2},
  pages = {023093},
  numpages = {19},
  year = {2022},
  month = {May},
  publisher = {American Physical Society},
  doi = {10.1103/PhysRevResearch.4.023093},
  url = {https://doi.org/10.1103/PhysRevResearch.4.023093}
}

@Article{osada.wang.20,
author={Osada, Motoki and Wang, Bai Yang and Goodge, Berit H.
and Lee, Kyuho and Yoon, Hyeok and Sakuma, Keita
and Li, Danfeng and Miura, Masashi and Kourkoutis, Lena F.
and Hwang, Harold Y.},
title={A Superconducting Praseodymium Nickelate with Infinite Layer Structure},
journal={Nano Lett.},
year={2020},
month={Aug},
day={12},
publisher={American Chemical Society},
volume={20},
number={8},
pages={5735-5740},
issn={1530-6984},
doi={10.1021/acs.nanolett.0c01392},
url={https://doi.org/10.1021/acs.nanolett.0c01392}
}

@article{osada.wang.21,
author = {Osada, Motoki and Wang, Bai Yang and Goodge, Berit H. and Harvey, Shannon P. and Lee, Kyuho and Li, Danfeng and Kourkoutis, Lena F. and Hwang, Harold Y.},
title = {Nickelate Superconductivity without Rare-Earth Magnetism: {(La,Sr)NiO$_{2}$}},
journal = {Adv. Mater.},
volume = {33},
number = {45},
pages = {2104083},
doi = {https://doi.org/10.1002/adma.202104083},
url = {https://onlinelibrary.wiley.com/doi/abs/10.1002/adma.202104083},
year = {2021}
}

@article{osada.wang.20b,
  title = {Phase diagram of infinite layer praseodymium nickelate {Pr$_{1-x}$Sr$_{x}$NiO$_{2}$} thin films},
  author = {Osada, Motoki and Wang, Bai Yang and Lee, Kyuho and Li, Danfeng and Hwang, Harold Y.},
  journal = {Phys. Rev. Mater.},
  volume = {4},
  issue = {12},
  pages = {121801},
  numpages = {5},
  year = {2020},
  month = {Dec},
  publisher = {American Physical Society},
  doi = {10.1103/PhysRevMaterials.4.121801},
  url = {https://doi.org/10.1103/PhysRevMaterials.4.121801}
}

@article{pardew.burke.96,
  title = {Generalized Gradient Approximation Made Simple},
  author = {Perdew, J. P. and Burke, K. and Ernzerhof, M.},
  journal = {Phys. Rev. Lett.},
  volume = {77},
  issue = {18},
  pages = {3865},
  numpages = {0},
  year = {1996},
  month = {Oct},
  publisher = {American Physical Society},
  doi = {10.1103/PhysRevLett.77.3865},
  url = {http://doi.org/10.1103/PhysRevLett.77.3865}
}

@Article{ptok.basak.23,
AUTHOR = {Ptok, Andrzej and Basak, Surajit and Piekarz, Przemysław and Oleś, Andrzej M.},
TITLE = {Influence of f Electrons on the Electronic Band Structure of Rare-Earth Nickelates},
JOURNAL = {Condensed Matter},
VOLUME = {8},
YEAR = {2023},
NUMBER = {1},
pages = {19},
URL = {https://doi.org/10.3390/condmat8010019},
ISSN = {2410-3896},
DOI = {10.3390/condmat8010019}
}

@Article{ren.li.23,
author={Ren, Xiaolin and Li, Jiarui
and Chen, Wei-Chih and Gao, Qiang
and Sanchez, Joshua J. and Hales, Jordyn
and Luo, Hailan and Rodolakis, Fanny
and McChesney, Jessica L. and Xiang, Tao
and Hu, Jiangping and Comin, Riccardo
and Wang, Yao and Zhou, Xingjiang
and Zhu, Zhihai},
title={Possible strain-induced enhancement of the superconducting onset transition temperature in infinite-layer nickelates},
journal={Commun. Phys.},
year={2023},
month={Nov},
day={27},
volume={6},
number={1},
pages={341},
issn={2399-3650},
doi={10.1038/s42005-023-01464-x},
url={https://doi.org/10.1038/s42005-023-01464-x}
}

@Article{rossi.osada.22,
author={Rossi, Matteo and Osada, Motoki and Choi, Jaewon
and Agrestini, Stefano and Jost, Daniel and Lee, Yonghun
and Lu, Haiyu and Wang, Bai Yang and Lee, Kyuho
and Nag, Abhishek and Chuang, Yi-De and Kuo, Cheng-Tai
and Lee, Sang-Jun and Moritz, Brian and Devereaux, Thomas P.
and Shen, Zhi-Xun and Lee, Jun-Sik and Zhou, Ke-Jin
and Hwang, Harold Y. and Lee, Wei-Sheng},
title={A broken translational symmetry state in an infinite-layer nickelate},
journal={Nat. Phys.},
year={2022},
month={Aug},
day={01},
volume={18},
number={8},
pages={869},
issn={1745-2481},
doi={10.1038/s41567-022-01660-6},
url={https://doi.org/10.1038/s41567-022-01660-6}
}

@article{ryee.yoon.20,
  title = {Induced magnetic two-dimensionality by hole doping in the superconducting infinite-layer nickelate {Nd$_{1-x}$Sr$_{x}$NiO$_{2}$}},
  author = {Ryee, Siheon and Yoon, Hongkee and Kim, Taek Jung and Jeong, Min Yong and Han, Myung Joon},
  journal = {Phys. Rev. B},
  volume = {101},
  issue = {6},
  pages = {064513},
  numpages = {5},
  year = {2020},
  month = {Feb},
  publisher = {American Physical Society},
  doi = {10.1103/PhysRevB.101.064513},
  url = {https://doi.org/10.1103/PhysRevB.101.064513}
}

@article{sakakibara.usui.20,
  title = {Model Construction and a Possibility of Cupratelike Pairing in a New ${d}^{9}$ Nickelate Superconductor {(Nd,Sr)NiO$_{2}$}},
  author = {Sakakibara, Hirofumi and Usui, Hidetomo and Suzuki, Katsuhiro and Kotani, Takao and Aoki, Hideo and Kuroki, Kazuhiko},
  journal = {Phys. Rev. Lett.},
  volume = {125},
  issue = {7},
  pages = {077003},
  numpages = {6},
  year = {2020},
  month = {Aug},
  publisher = {American Physical Society},
  doi = {10.1103/PhysRevLett.125.077003},
  url = {https://doi.org/10.1103/PhysRevLett.125.077003}
}

@Article{tam.choi.22,
author={Tam, Charles C.
and Choi, Jaewon
and Ding, Xiang
and Agrestini, Stefano
and Nag, Abhishek
and Wu, Mei
and Huang, Bing
and Luo, Huiqian
and Gao, Peng
and Garc{\'i}a-Fern{\'a}ndez, Mirian
and Qiao, Liang
and Zhou, Ke-Jin},
title={Charge density waves in infinite-layer {NdNiO$_{2}$} nickelates},
journal={Nat. Mater.},
year={2022},
month={Oct},
day={01},
volume={21},
number={10},
pages={1116},
issn={1476-4660},
doi={10.1038/s41563-022-01330-1},
url={https://doi.org/10.1038/s41563-022-01330-1}
}

\end{document}